%% file: eprint.tex
\documentclass[12pt]{article}
\usepackage{graphicx}

\input{babarsym}

\input{symbols}

\newcommand\pubnumber{}
\newcommand\pubdate{\today}

\def\warwick{Department of Physics\\
University of Warwick, Coventry, CV4 7AL, UK}
\def\support{\footnote{Speaker on behalf of the \babar\ Collaboration}}

\def\Title#1{\begin{center} {\Large #1 } \end{center}}
\def\Author#1{\begin{center}{ \sc #1} \end{center}}
\def\Address#1{\begin{center}{ \it #1} \end{center}}

\newcommand\pubblock{\rightline{\begin{tabular}{l} \pubnumber\\
         \pubdate  \end{tabular}}}
\newenvironment{Abstract}{\begin{quotation}  }{\end{quotation}}
\newenvironment{Presented}{\begin{quotation} \begin{center} 
             PROCEEDINGS OF\end{center}\bigskip 
      \begin{center}\begin{large}}{\end{large}\end{center} \end{quotation}}
\def\Acknowledgements{\bigskip  \bigskip \begin{center} \begin{large}
             \bf ACKNOWLEDGEMENTS \end{large}\end{center}}
\input econfmacros.tex
\begin{document}
\begin{titlepage}
\pubblock

\vfill
\Title{Determination of $\gamma$ from $B\to K^{*}\pi$ decays and related modes}
\vfill
\Author{Eugenia Maria Teresa Irene Puccio\support}
\Address{\warwick}
\vfill
\begin{Abstract}
We present the status of recent results from the \babar\ and Belle
experiments on the measurement of the angle $\gamma$ from the Dalitz plot
analyses of \BztoKspipi\ and \BztoKppimpiz. 
\end{Abstract}
\vfill
\begin{Presented}
CKM2010, the 6th International Workshop on the CKM Unitarity Triangle\\
University of Warwick, UK,  September 6--10, 2010
\end{Presented}
\vfill
\end{titlepage}
\def\thefootnote{\fnsymbol{footnote}}
\setcounter{footnote}{0}

\section{Introduction}

At tree level, $B\to K^{*}\pi$ decays are sensitive to $\gamma$ through the
relation
\begin{equation}
e^{-2i\gamma}\propto\frac{\bar{A}\left(K^{*-}\pi^{+}\right)+\sqrt{2}\bar{A}\left(\bar{K}^{*0}\pi^{0}\right)}{A\left(K^{*+}\pi^{-}\right)+\sqrt{2}A\left(K^{*0}\pi^{0}\right)},
\end{equation}
where $A$ and $\bar{A}$ are the decay amplitude and its charge conjugate
respectively. To measure $\gamma$, three-body decays have an advantage over
quasi-two body decays since $B\to K^{*}\pi$ can interfere through
the same final state in $B\to K\pi\pi$. By measuring the interference
pattern in the Dalitz plot, it is possible to determine not only magnitudes
of the amplitudes as in the two body decays but also the relative phases
between the amplitudes. The cleanest method to determine $\gamma$ from
$K\pi\pi$ Dalitz plots involves the charmless decays \BztoKppimpiz\ and
\BztoKspipi~\cite{Ciuchini:2006kv,Gronau:2006qn}.
The method involves forming isospin triangles from $K^{*}\pi$ intermediate
modes in \BztoKppimpiz\ and \BztoKspipi.  By using isospin decomposition,
the QCD penguin contributions in $B\to K^{*}\pi$ decays are cancelled and
the resultant amplitude is as follows:
\begin{equation}
3A_{3/2}=A\left(\BztoKstarppim\right)+\sqrt{2}A\left(\BztoKstarzpiz\right),
\label{resultant-A}
\end{equation}
with an equivalent amplitude for the charge-conjugate state,
$\bar{A}_{3/2}$. In the absence of electroweak penguins(EWP), $A_{3/2}$
carries a weak phase $\gamma$ so that in this limit
\begin{equation}
\gamma=\Phi_{3/2}=-\frac{1}{2}\arg\left(\frac{\bar{A}_{3/2}}{A_{3/2}}\right).
\label{resultant-P}
\end{equation}
The phase $\Phi_{3/2}$ can be determined by measuring the
following quantities:
\begin{itemize}
\item phase $\Delta\phi$, between \BztoKstarppim\ and \BzbtoKstarmpip\ in
\BztoKspipi. 
\item phase $\phi$, between \BztoKstarppim\ and \BztoKstarzpiz\ in
\BztoKppimpiz; 
\item its charge conjugate equivalent in \BzbtoKmpip\piz; 
\end{itemize}
This method to extract $\gamma$ is similar to the Snyder-Quinn method used
to obtain $\alpha$ from \Bztopipipiz~\cite{Snyder:1993mx}.  $B\to\rho\pi$
amplitudes, measured from the three body decay of \Bztopipipiz, are used in
this method to provide an SU(3) correction for EWP contributions, necessary
to obtain a constraint for $\gamma$. 

\section{Experimental Results: $\Delta\phi$}

\begin{figure}
\begin{tabular}{lll}
\includegraphics[width=0.31\textwidth]{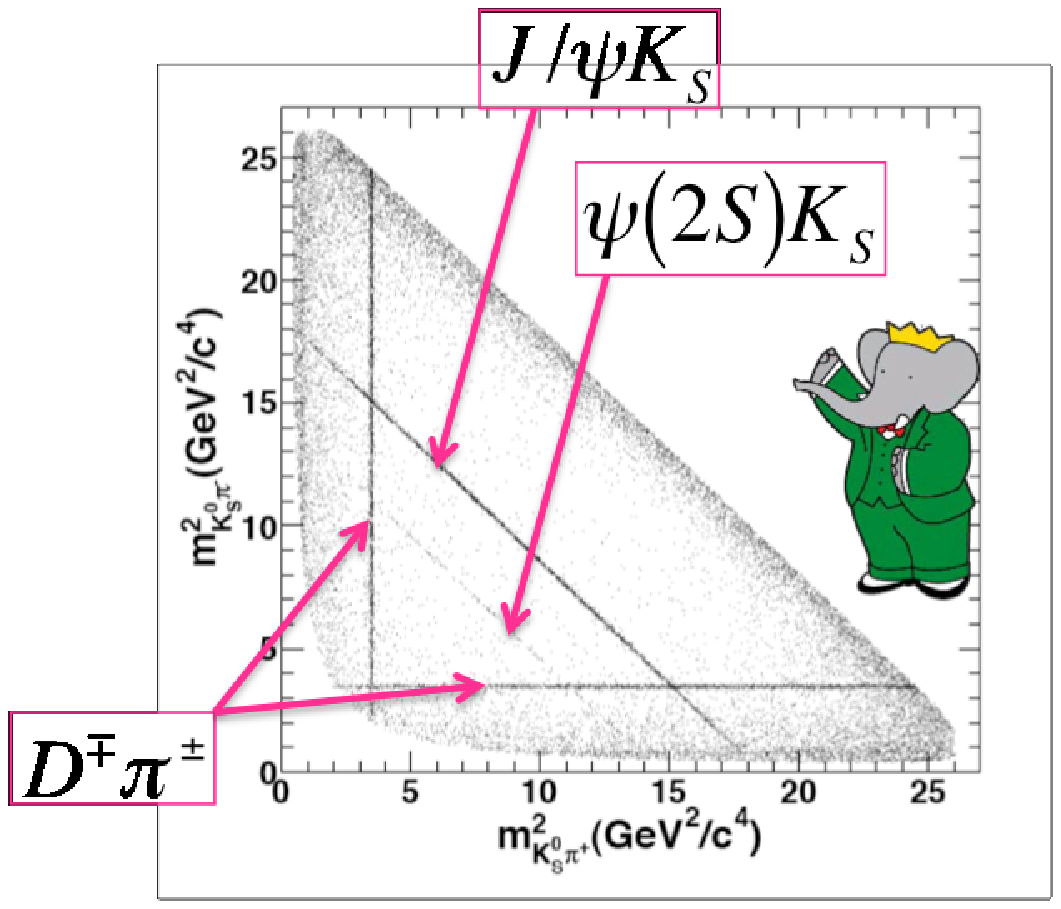} &
\includegraphics[width=0.35\textwidth]{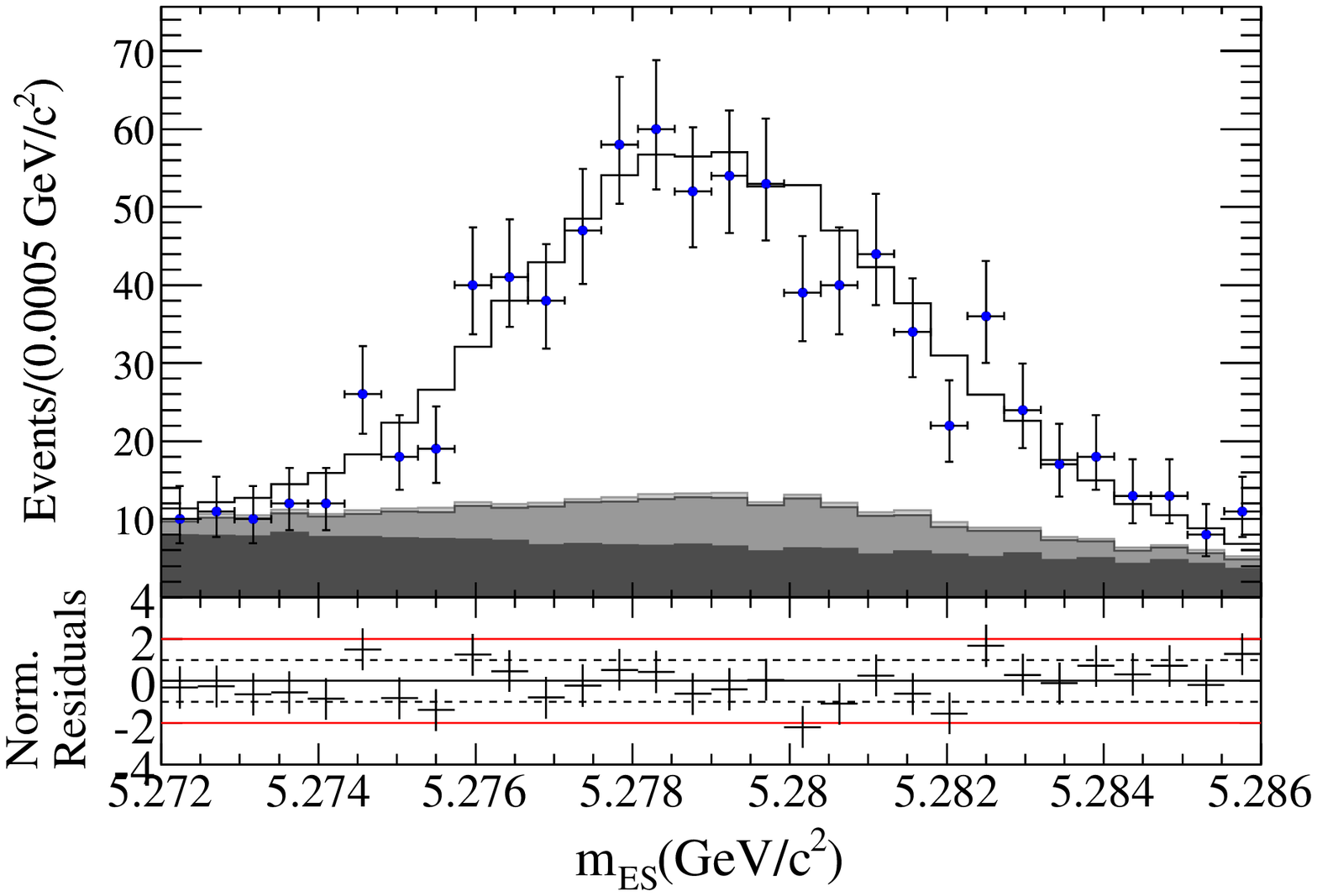} &
\includegraphics[width=0.35\textwidth]{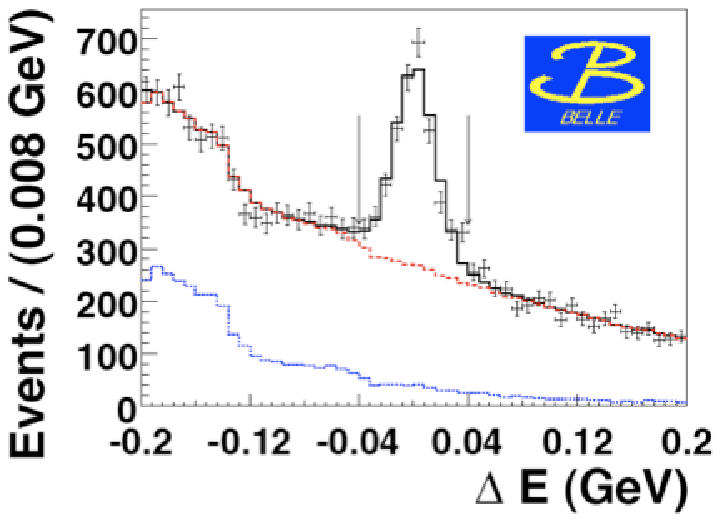} 
\end{tabular}
\caption{Resultant Dalitz plot distribution (left) and projection plots for
\mes\ (center, \babar\ results)~\cite{Aubert:2009me} and \DeltaE\ (right,
Belle result)~\cite{:2008wwa}.}
\label{fig:DP-proj}
\end{figure}

\begin{figure}[htb]
\centering
\begin{tabular}{lll}
\includegraphics[width=0.32\textwidth]{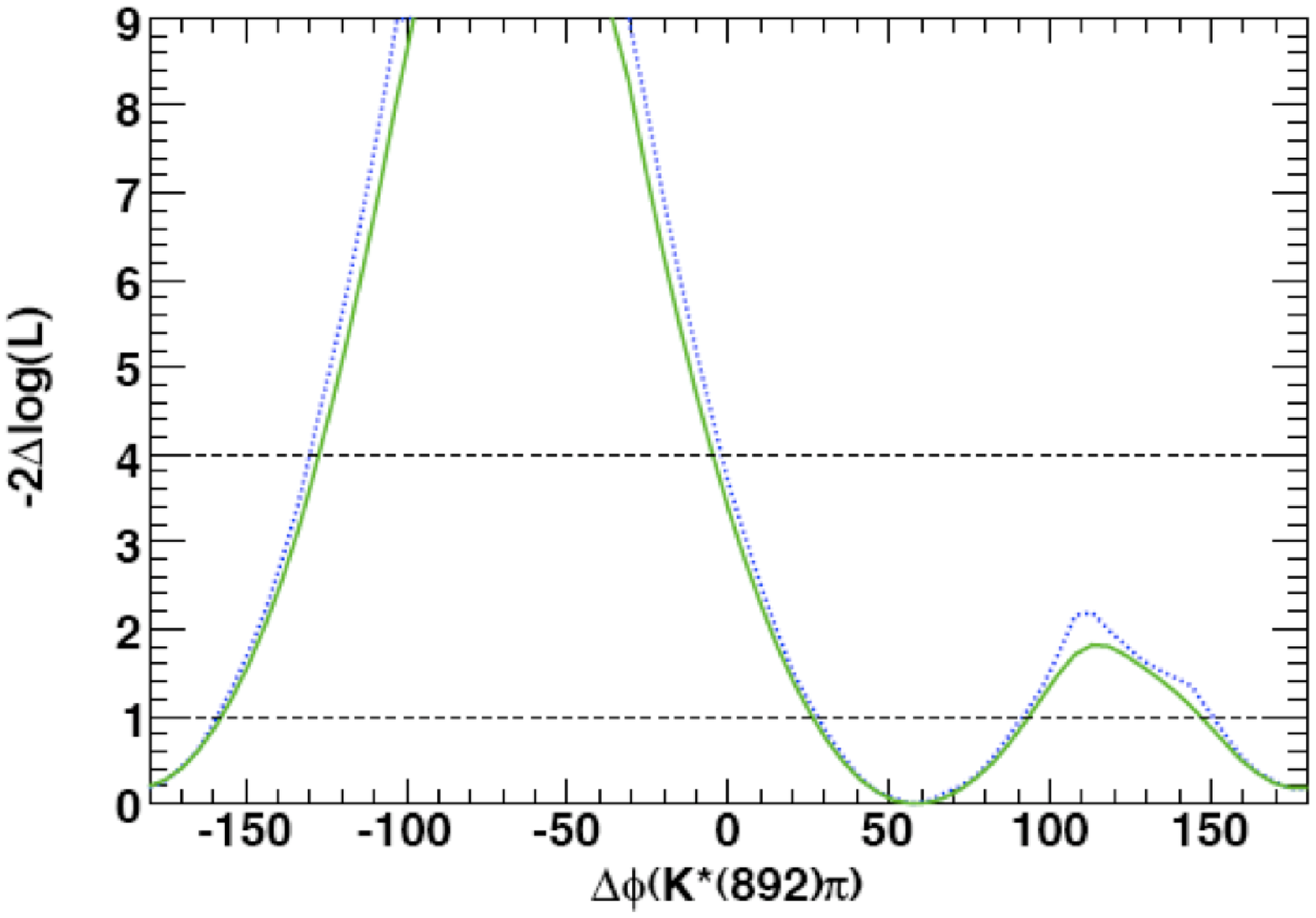} &
\includegraphics[width=0.32\textwidth]{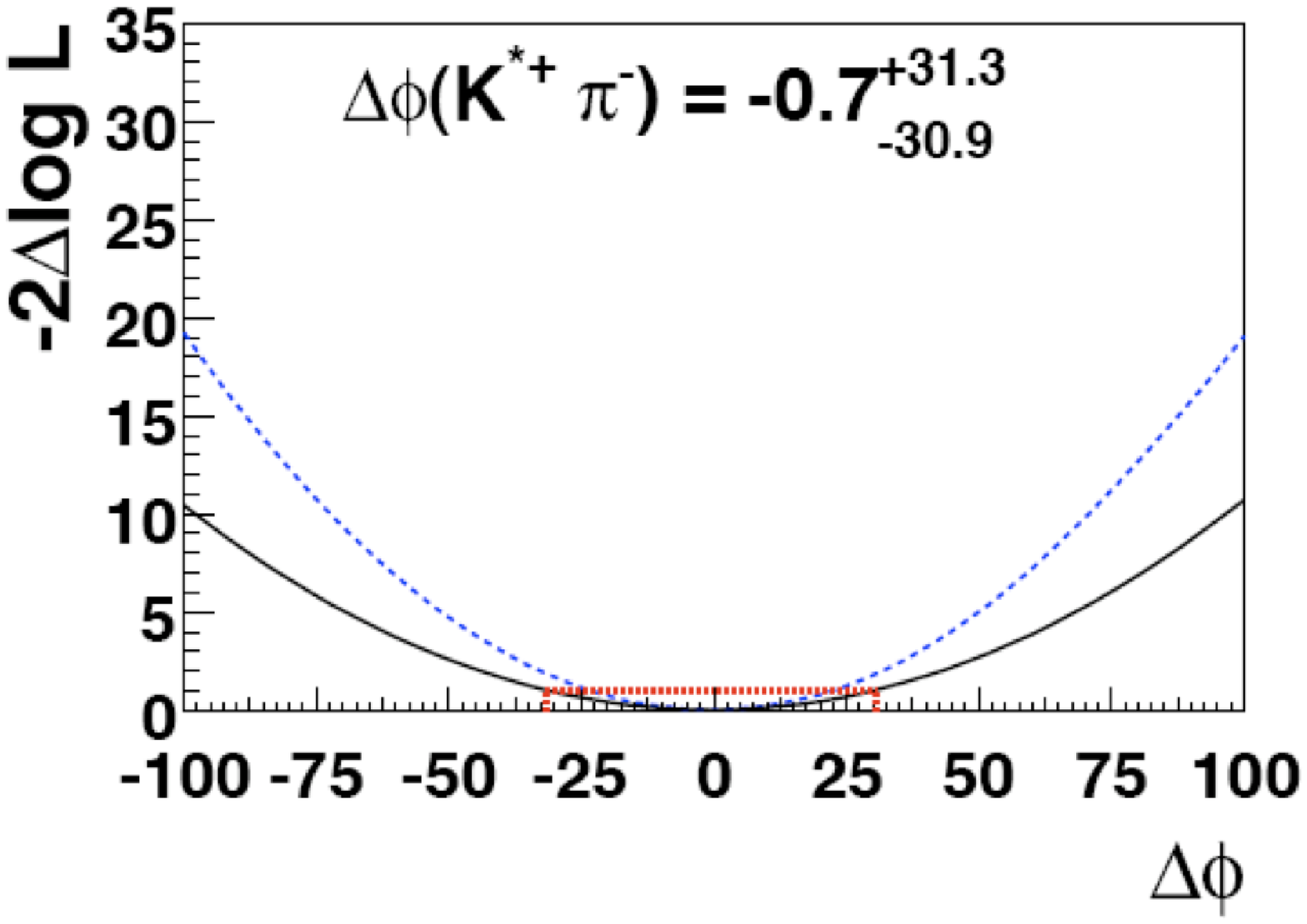} &
\includegraphics[width=0.32\textwidth]{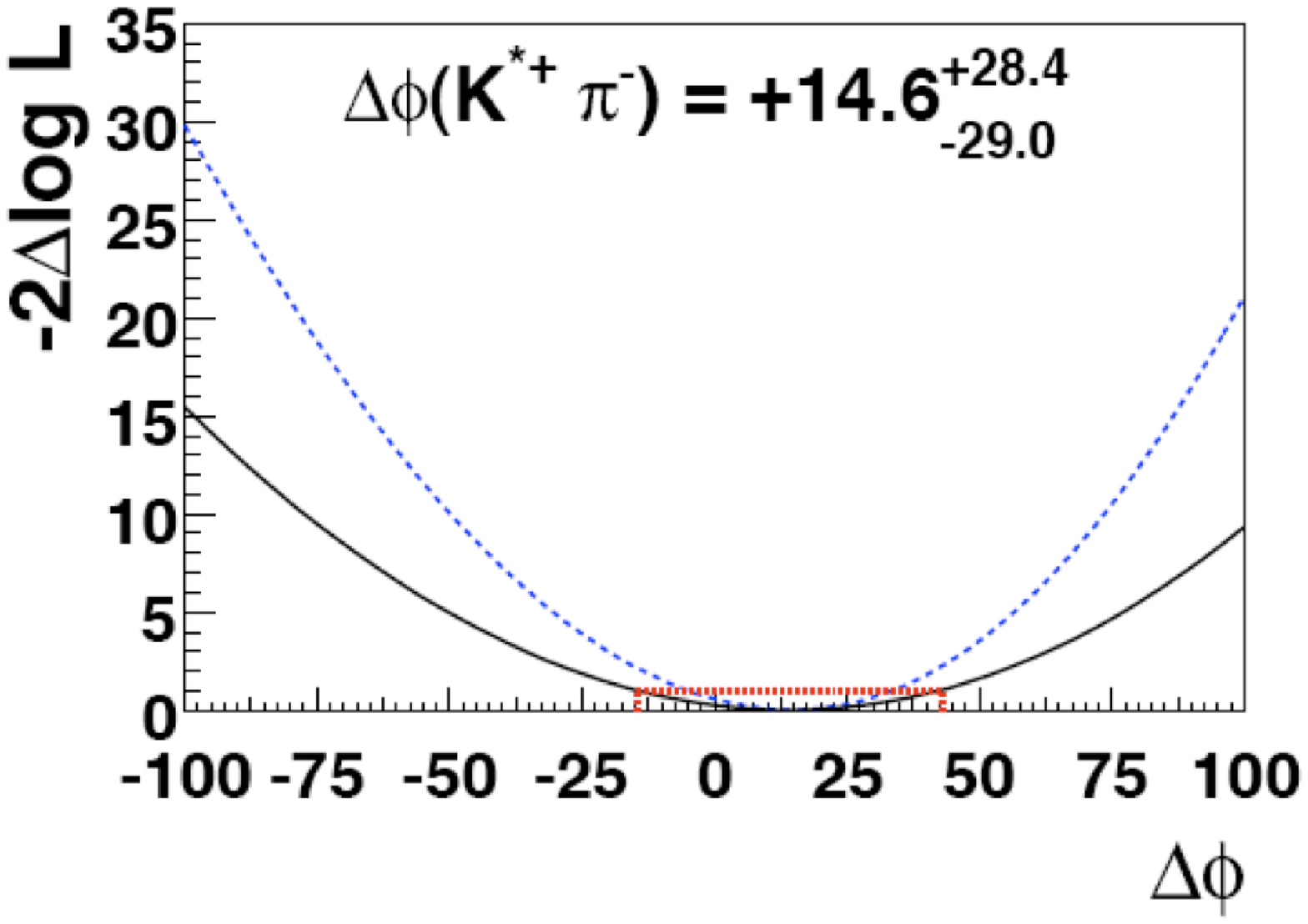}
\end{tabular}
\caption{Likelihood scans of $\Delta\phi$ from Dalitz plot analyses of
\BztoKspipi. The left likelihood distribution is taken from the \babar\
results~\cite{Aubert:2009me}, the centre and right distributions are from
Belle~\cite{:2008wwa} and represent the scans of the two different
solutions.}
\label{fig:Dphi-kspipi}
\end{figure}

The \BztoKspipi\ Dalitz plot provides the phase difference $\Delta\phi$
between \BztoKstarppim\ and \BzbtoKstarmpip measured from
$\Delta\phi_{K^{*}\pi}=\bar{\phi}_{\Kstarm\pip}-\phi_{\Kstarp\pim}$
To obtain $\Delta\phi$, the $K^{*}\pi$ phases need to be
measured relative to each other, taking into account also the additional
phase of $-2\beta$.  The relative phases are determined at the interference
regions around the edges of the Dalitz plot. However the overlap region of
resonances is small and the effect on event density small, making it
crucial to understand backgrounds and efficiencies in the interference
regions.  The main background contribution in this Dalitz plot is found to
come from continuum events and those are mostly rejected by a Neural
Network. The remaining background contribution are \B\ meson decays to
charm final states, shown as bands in the resultant Dalitz plot
distribution in \figref{DP-proj}. Projection plots for signal and
background of discriminating variables, \mes\ taken from the \babar\ result
of $383$ million \BB\ events~\cite{Aubert:2009me} and \DeltaE\ from the
Belle result of $657$ million \BB\ events~\cite{:2008wwa}, are also shown
in \figref{DP-proj}.
The results of the likelihood scans for $\Delta\phi$ are shown in
\figref{Dphi-kspipi} and summarised in \tabref{deltaPhi-res}.
Two fit solutions are found corresponding to the interference between
\KstarIII\ and the nonresonant component. These two solutions give
different results for the values of $\Delta\phi$. There is some
disagreement between the \babar\ and Belle results. The experimentally
measured values of $\Delta\phi$ shown in \tabref{deltaPhi-res} include the
\BzBzb\ mixing phase and this has to be removed before the values can be
used in the extraction of $\gamma$.

\begin{table}[t]
\begin{center}
\begin{tabular}{lc}  
\hline
Experiment	& $\Delta\phi\left(\Kstarp\pim\right)$ \\
\hline
\babar\ Soln. 1	& $\left(58.3\pm32.7\pm4.6\pm8.1\right)^{\circ}$    \\
\babar\ Soln. 2	& $\left(176.6\pm28.8\pm4.6\pm8.1\right)^{\circ}$    \\
Belle Soln. 1	& $\left(-0.7\pm^{24}_{23}\pm11\pm18\right)^{\circ}$ \\
Belle Soln. 1	& $\left(14.6\pm^{19}_{20}\pm11\pm18\right)^{\circ}$ \\
\hline
\end{tabular}
\caption{Summary of the results for $\Delta\phi\left(\Kstarp\pim\right)$
from time-dependent Dalitz plot analyses of \BztoKspipi. The uncertainties
quoted are statistical, systematic and model-dependent respectively.}
\label{tab:deltaPhi-res}
\end{center}
\end{table}

\section{Experimental Results: $\phi$ and $\bar{\phi}$}

The other two parameters required to determine $\gamma$ are $\phi$ and its
charge conjugate, $\bar{\phi}$. These are the relative phases between
\BztoKstarppim\ and \BztoKstarzpiz\ and \BzbtoKstarmpip\ and
\BzbtoKstarzbpiz\ respectively:
\begin{eqnarray}
\phi = \phi_{\Kstarz\piz}-\phi_{\Kstarp\pim}	& &
\bar{\phi} = \bar{\phi}_{\Kstarzb\piz}-\bar{\phi}_{\Kstarm\pip} 
\end{eqnarray}
Both of these relative phases are determined from Dalitz plot analysis of
\BztoKppimpiz\ and its charge conjugate.  Preliminary results are available
from the full \babar\ dataset of $454$ million \BB\
events~\cite{Wagner:Thesis}.  Expanding
Eq.~\ref{resultant-P}, $\Phi_{3/2}$ is obtained from the combination of the
phases $\phi$ and $\bar{\phi}$ and subtracting the phase $\Delta\phi$
obtained from the time-dependent Dalitz plot analysis of \BztoKspipi.  

\section{Issues with interpretation}

The choice of the phase convention is important when combining the results
since failure to take the convention into account can result in a
$180^{\circ}$ shift in relative phase~\cite{Gronau:2010kq}. The amplitude
is proportional to the cosine of a helicity angle between the final states
particles in a three-body decay. The helicity convention defines an
ordering of particles in the SU(2) decomposition that can introduce a sign
flip in Eq.~\ref{resultant-A}. Therefore relative phases between vector
amplitudes need to be interpreted with respect to a given helicity
convention. Another issue
with the interpretation of the results is that whereas QCD penguin
contributions cancel in the sum of $A_{K^{*}\pi}$ so that
Eq.~\ref{resultant-A} is QCD penguin free, EW penguin contributions still
need to be accounted for.  SU(3) decomposition of operators gives a good
approximation to
\begin{eqnarray}
A_{3/2} = Te^{i\gamma}-P_{EWP} & &
A_{3/2}	\propto
	\left(\bar{\rho}+i\bar{\eta}\right)\left(1+r_{3/2}\right)+C\left(1-r_{3/2}\right),
	\label{su3-decom} 
\end{eqnarray}
where $T$ and $P_{EWP}$ are the tree and EW penguin contributions
respectively. $C$ in Eq.~\ref{su3-decom} depends only on EW physics and is
well known to a theoretical error below $1\%$ with $C=-0.27$. The quantity
$r_{3/2}$ is the ratio of hadronic matrix elements and is measured
from~\cite{Gronau:2006qn,Antonelli:2009ws}:
\begin{equation}
r_{3/2}=\frac{\left[A_{\rhop\piz}-A_{\rhoz\pip}\right]-\sqrt{2}\left[A_{\Kstarp\Kzb}-A_{\Kp\Kstarzb}\right]}{A_{\rhop\piz}+A_{\rhoz\pip}}.
\end{equation}
Current experimental results for these quantities are shown in
\tabref{ratio-decays}. $B\to\rho\pi$ decays have well known BFs and
$A_{\CP}$, however amplitudes for $KK^{*}$ decays are small but the
relative phases are unknown. The strategy used is to separate the ratio
into well-measured components, add the $KK^{*}$ ratio as a systematic
uncertainty and account for $m_{s}/\Lambda_{QCD}\approx30\%$ of SU(3)
breaking. Preliminary results for $r_{3/2}$ and subsequently for the EW
penguin to tree amplitude ratio are~\cite{Wagner:Thesis}
\begin{eqnarray}
Re(r_{3/2})       &=& 0.21\pm0.13(stat.)\pm0.77(syst.)\pm0.06(theo.), \\
\pm Im(r_{3/2})   &=& 1.45\pm0.35(stat.)\pm0.77(syst.)\pm0.44(theo.), \\
Re(P_{EWP}/T)	  &=& -0.21\pm0.13(stat.)\pm0.29(syst.)\pm0.16(theo.), \\
\pm Im(P_{EWP}/T) &=& -0.54\pm0.05(stat.)\pm0.29(syst.)\pm0.04(theo.).
\end{eqnarray}
The systematic uncertainty is the dominant source of error in this
measurement and can only be eliminated by measuring the relative phases for
$\Kstarp\Kzb$ and $\Kp\Kstarzb$.
\begin{table}[t]
\begin{center}
\begin{tabular}{lcc}  
\hline
Decay model	& BF $(\times 10^{-6})$	& $A_{\CP}$ \\
\hline
\Bptorhozpip 	& $8.3^{+1.2}_{-1.3}$	& $0.18^{+0.09}_{-0.17}$	\\
\Bptorhoppiz	& $10.9^{+1.4}_{-1.5}$	& $0.02\pm0.11$			\\
\BptoKpKstarzb	& $0.68\pm0.19$		& $ - $ 			\\
\BptoKsKspip	& $<0.51$		& $ - $ 			\\
\hline
\end{tabular}
\caption{Current experimental results for BF and $A_{\CP}$ for two body and
quasi-two body $\rho\pi$ and $K^{*}K$ decays as taken from HFAG Winter
2010~\cite{TheHeavyFlavorAveragingGroup:2010qj}.}
\label{tab:ratio-decays}
\end{center}
\end{table}

\section{Conclusion}

\babar\ results for \Kppimpiz\ are in process of being finalised and
results should soon be combined to form the CKM constraint.  The angle
$\gamma$ can also be measured by looking at the phase difference from $\rho
K$ and $K^{*}\pi$. Tree to QCD penguin ratio is expected to be larger in
$\rho K$ than in $K^{*}\pi$ giving a potentially better sensitivity to
$\gamma$.  This method is also quite promising for future experiments. A
Super B factory can expect results with uncertainties a factor $\sim15$
smaller than \babar's.  LHCb could also have potential for these
measurements and additionally study the constraint in the \Bs\
decays~\cite{Ciuchini:2006st}.

\Acknowledgements

I am grateful to the organisers and participants of CKM 2010 for this
opportunity to present and discuss these results. I would like to also
thank my colleagues on \babar\ and Belle who have made these measurements
possible. In particular I would like to thank Tim Gershon, Thomas Latham,
Mathew Graham and Andrew Wagner for their help and support in preparing
this presentation.

\end{document}

%% file: symbols.tex
\newcommand{\eqref}[1]{Eq.~(\ref{eq:#1})}

\newcommand{\Kstarzpiz}          {\mbox{$\Kstarz \piz$}}
\newcommand{\BztoKstarppim}     {\mbox{$\Bz \to \Kstarp \pim$}}
\newcommand{\BzbtoKstarmpip}     {\mbox{$\Bzb \to \Kstarm \pip$}}
\newcommand{\BztoKstarzpiz}     {\mbox{$\Bz \to \Kstarzpiz$}}
\newcommand{\BzbtoKstarzbpiz}     {\mbox{$\Bzb \to \Kstarzb\piz$}}
\newcommand{\BptoKpKstarzb}     {\mbox{$\Bp \to \Kp\Kstarzb$}}

\newcommand{\Kspipi}             {\mbox{$\KS\pip\pim$}}

\newcommand{\BztoKspipi}         {\mbox{$\Bz \to \Kspipi$}}

\newcommand{\KstarIII}           {\mbox{$\Kstarz_{2}(1430)$}}

\newcommand{\rhoz}               {\mbox{$\rho^0$}}
\newcommand{\rhop}               {\mbox{$\rho^+$}}

\newcommand{\Bptorhoppiz}         {\mbox{$\Bp \to \rhop \piz$}}
\newcommand{\Bptorhozpip}         {\mbox{$\Bp \to \rhoz \pip$}}

\newcommand{\BptoKsKspip}          {\mbox{$\Bp \to \KS\KS\pip$}}

\newcommand{\Bztopipipiz}        {\mbox{$\Bz \to \pip\pim\piz$}}

\newcommand{\BztoKppimpiz}          {\mbox{$\Bz \to \Kp \pim \piz$}}
\newcommand{\BzbtoKmpip}         {\mbox{$\Bzb \to \Km \pip$}}

\newcommand{\Kppimpiz}       {\mbox{$\Kp\pim\piz$}}

%% file: econfmacros.tex
\def\beq{\begin{equation}}
\def\eeq#1{\label{#1}\end{equation}}
\def\eeqn{\end{equation}}

\def\beqa{\begin{eqnarray}}
\def\eeqa#1{\label{#1}\end{eqnarray}}
\def\eeqan{\end{eqnarray}}

\let\bar=\overbar

\def\Dslash{\not{\hbox{\kern-4pt $D$}}}
\def\dslash{\not{\hbox{\kern-2pt $\del$}}}

\def\BR{\mbox{\rm BR}}

\def\msb{{\bar{\ssstyle M \kern -1pt S}}}